\documentclass[fleqn,12pt,twoside]{article}
\usepackage[headings]{espcrc1}

\usepackage{lineno}

\readRCS
$Id: espcrc1.tex,v 1.2 2004/02/24 11:22:11 spepping Exp $
\usepackage{graphicx}
\usepackage[figuresright]{rotating}

\newcommand{\AmS}{{\protect\the\textfont2
  A\kern-.1667em\lower.5ex\hbox{M}\kern-.125emS}}

\begin{document}

\title{Spectral modeling of scintillator for the NEMO-3 and SuperNEMO detectors}


\author{J.~Argyriades \address[LAL] {LAL, Universit\'e Paris-Sud, CNRS/IN2P3, F-91405 Orsay, France},
R.~Arnold \address[Strasbourg] {IPHC, Universit\'e de Strasbourg, CNRS/IN2P3, F-67037, Strasbourg, France},
C.~Augier \addressmark[LAL],
J.~Baker \address[INEL] {INL, Idaho Falls, 83415, USA},
A.S.~Barabash \address[ITEP] {Institute of Theoretical and Experimental Physics, 117259 Moscow, Russia},
M.~Bongrand \addressmark[LAL],
G.~Broudin-Bay \addressmark[LAL],
V.B.~Brudanin \address [JINR] {Joint Institute for Nuclear Research, 141980 Dubna, Russia},
A.J.~Caffrey \addressmark[INEL],
S.~Cebri\'an \address[Zaragoza]{University of Zaragoza, C/ Pedro Cerbuna 12, 50009 Zaragoza, Spain},
A.~Chapon \address [LPC] {LPC Caen, ENSICAEN, Universit\'e de Caen, CNRS/IN2P3, F-14032 Caen, France},
E.~Chauveau \address [CENBG]{CNRS/IN2P3, Centre d'Etudes Nucl\'eaires de Bordeaux Gradignan, UMR 5797, F-33175 Gradignan, France}\address [UB]{Universit\'e de Bordeaux, Centre d'Etudes Nucl\'eaires de Bordeaux Gradignan, UMR 5797, F-33175 Gradignan, France},
Th.~Dafni \addressmark[Zaragoza],
Z.~Daraktchieva \address [UCL] {University College London, WC1E 6BT London, United Kingdom},
J.~D\'{\i}az \address [IFIC] {IFIC, CSIC - Universidad de Valencia, Valencia, Spain},
D.~Durand \addressmark[LPC],
V.G.~Egorov \addressmark[JINR],
J.J.~Evans \addressmark [UCL],
N.~Fatemi-Ghomi \address[UM] {University of Manchester, M13 9PL Manchester, United Kingdom},
R.~Flack \addressmark[UCL],
A.~Basharina-Freshville \addressmark[UCL],
K-I.~Fushimi \address[Tokushima]{Tokushima University, 770-8502, Japan},
X.~Garrido \addressmark[LAL],
H.~G\'omez \addressmark[Zaragoza],
B.~Guillon \addressmark[LPC],
A.~Holin \addressmark[UCL],
K.~Hol\'{y} \address[Comenius]{Comenius University, Department of Nuclear Physics and Biophysics, Mlynsk\'{a} dolina, SK-84248, Bratislava, Slovakia},
J.J.~Horkley \addressmark[INEL],
Ph.~Hubert  \addressmark[CENBG]\addressmark[UB],
C.~Hugon \addressmark[CENBG],
F.J.~Iguaz \addressmark[Zaragoza],
I.G.~Irastorza \addressmark[Zaragoza],
N.~Ishihara \address[KEK]{KEK, 1-1 Oho, Tsukuba, Ibaraki 305-0801, Japan},
C.M.~Jackson \addressmark[UM],
S.~Jullian \addressmark[LAL],
S.~Kanamaru \address[Osaka]{Osaka University, 1-1 Machikaneyama Toyonaka, Osaka 560-0043, Japan},
M.~Kauer \addressmark[UCL],
O.I.~Kochetov \addressmark[JINR],
S.I.~Konovalov \addressmark[ITEP],
V.E.~Kovalenko \addressmark[JINR]\addressmark[Strasbourg],
D.~Lalanne \addressmark[LAL],
K.~Lang \address[Texas]{Department of Physics, 1 University Station C1600, The University of Texas at Austin, Austin, TX 78712, USA},
Y.~Lemi\`{e}re \addressmark[LPC],
G.~Lutter \addressmark[CENBG]\addressmark[UB],
G.~Luz\'on \addressmark[Zaragoza],
F.~Mamedov \address [CTU] {IEAP, Czech Technical University in Prague, CZ-12800 Prague, Czech Republic},
Ch.~Marquet \addressmark[CENBG]\addressmark[UB],
J.~Martin-Albo \addressmark[IFIC],
F.~Mauger \addressmark[LPC],
F.~Monrabal \addressmark[IFIC],
A.~Nachab \addressmark[CENBG]\addressmark[UB],
I.~Nasteva \addressmark[UM],
I.B.~Nemchenok \addressmark[JINR],
C.H.~Nguyen \addressmark[CENBG]\addressmark[UB],
F.~Nova \address [Barc] {Universitat Aut\`onoma de Barcelona, Spain},
P.~Novella \addressmark[IFIC],
H.~Ohsumi \address [Saga] {Saga University, Saga 840-8502, Japan},
R.~B.~Pahlka \addressmark[Texas]  \thanks{Corresponding author:  pahlka@hep.utexas.edu (R.~B.~ Pahlka) },
F.~Perrot \addressmark[CENBG]\addressmark[UB],
F.~Piquemal \addressmark[CENBG]\addressmark[UB],
P.P.~Povinec \addressmark[Comenius],
B.~Richards \addressmark [UCL],
J.S.~Ricol \addressmark[CENBG]\addressmark[UB],
C.L.~Riddle \addressmark[INEL],
A.~Rodriguez \addressmark[Zaragoza],
R.~Saakyan \addressmark[UCL],
X.~Sarazin \addressmark[LAL],
J.K.~Sedgbeer \address [Imperial] {Imperial College London, SW7 2AZ, London, United Kingdom},
L.~Serra  \addressmark[IFIC],
L.~Simard \addressmark[LAL],
F.~\v{S}imkovic \addressmark[Comenius],
Yu.A.~Shitov \addressmark[JINR]\addressmark[Imperial],
A.A.~Smolnikov \addressmark[JINR],
S.~S\"{o}ldner-Rembold \addressmark[UM],
I.~\v{S}tekl \addressmark[CTU],
Y.~Sugaya \addressmark[Osaka],
C.S.~Sutton \address [MHC] {MHC, South Hadley, Massachusetts, 01075, USA},
G.~Szklarz \addressmark[LAL],
Y.~Tamagawa \address[Fukui]{Fukui University, 6-10-1 Hakozaki, Higashi-ku, Fukuoka 812-8581, Japan},
J.~Thomas \addressmark[UCL],
R.~Thompson \addressmark[UM],
V.V.~Timkin \addressmark[JINR],
V.I.~Tretyak \addressmark[JINR]\addressmark[IPHC],
Vl.I.~Tretyak \address [KINR] {INR, MSP 03680 Kyiv, Ukraine},
V.I.~Umatov \addressmark[ITEP],
L.~V\'{a}la \addressmark[CTU],
I.A.~Vanyushin \addressmark[ITEP],
R.~Vasiliev \addressmark[JINR],
V.~Vorobel \address [Charles] {Charles University in Prague, Faculty of Mathematics and Physics, CZ-12116 Prague, Czech Republic},
Ts.~Vylov \addressmark[JINR],
D.~Waters \addressmark[UCL],
N.~Yahlali \addressmark[IFIC],
A.~\v{Z}ukauskas \addressmark [Charles]
}

%
%


\runtitle{1-column format camera-ready paper in \LaTeX}
\runauthor{B. Pahlka}



\maketitle

\begin{abstract}

We have constructed a GEANT4-based detailed software model of photon transport in plastic scintillator blocks and have used it to study the NEMO-3 and SuperNEMO calorimeters employed in experiments designed to search for neutrinoless double beta decay.  We compare our simulations to measurements using conversion electrons from a calibration source of $\rm ^{207}Bi$  and show that the agreement is improved if wavelength-dependent properties of the calorimeter are taken into account. In this article, we briefly describe our modeling approach and results of our studies.

\end{abstract}

\section{Introduction}

Several dedicated efforts have recently been proposed to describe optical photon transport in scintillator detectors using various Monte Carlo packages~\cite{Frlez:2000aw,Ghal-Eh,Cayouette:2003,Tickner:2007,Senchishin:2006qw}.  Using the GEANT4 (version 4.9.1 patch 3) framework~\cite{Agostinelli:2002hh}, we have constructed a comprehensive and detailed model of photon transport in plastic scintillator, then used the model to study the individual NEMO-3 calorimeter modules. In the model, we account for the wavelength dependence of optical properties of the scintillators, light guides, reflective wrappings, photodetectors and coupling materials.  We use wavelength dependent self-absorption and re-emission in the scintillator and account for the fluorescent quantum yield of the wavelength shifter.  Our results show that this detailed modeling exhibits better agreement with measurements compared to a monochromatic approach.

The NEMO-3 experiment, located at the Laboratoire Souterrain de Modane in the Fr\'ejus tunnel, searches for neutrinoless double beta decay by employing tracking and calorimetry systems and has been taking data since 2003~\cite{Arnold:2004xq,Arnold:2005rz,Argyriades:2009vq,Arnold:2006fk,Arnold:2006sd}.  The calorimeter modules consist of large polystyrene scintillator blocks with light guides coupled to either flat or hemispherical photomultiplier tubes (PMTs).  Signals in an individual block are due to incident particles, mostly $\beta$ and $\gamma$ rays, and the response varies with the energy and the impact point on the entrance face. The response also depends on the size and geometry of the blocks.  The energy resolution and background rejection improves if a correction for the non-uniformity due to the impact position is applied for each type of employed blocks~\cite{Arnold:2004xq}.  We have reproduced the spatial dependence of response of the NEMO-3 scintillators to $\rm ^{207}Bi$ conversion electrons and have optimized the new scintillator block geometry for the next generation double beta decay experiment, SuperNEMO.  

\section{Modeling details}

\subsection{The detector}

\begin{figure}[t]
\begin{center}
\includegraphics[width=30pc]{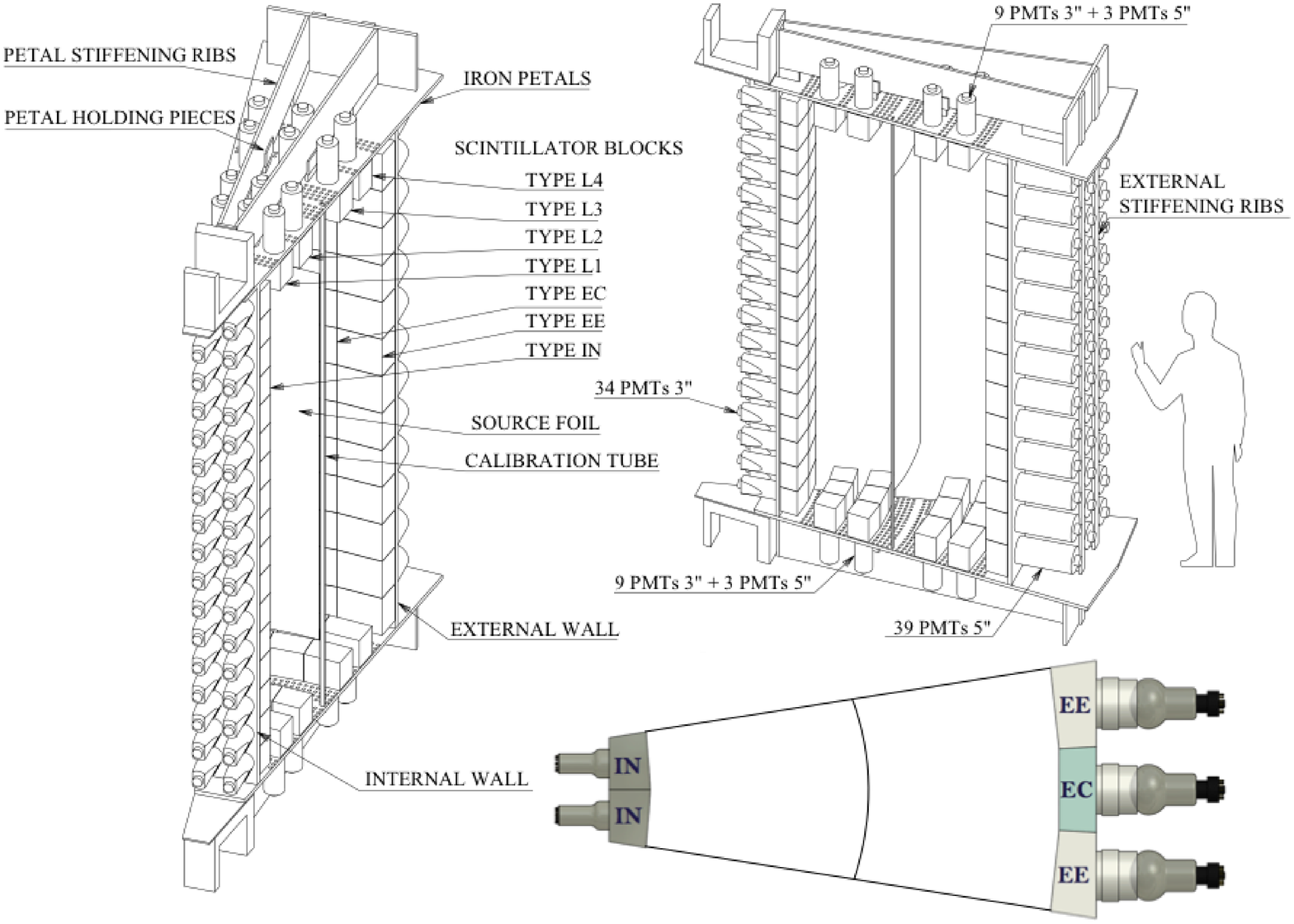}

\caption[NEMO-3 Blocks]{One of twenty sectors of the NEMO-3 detector with details showing the source foil, scintillator blocks, and photomultipliers.  EE, EC, and IN identify blocks on the exterior and interior walls.  L1 - L4 identify blocks on the petals (not modeled in this work). The lower figure shows a 2-D rendering of the wall blocks.}
\label{figure:blocks3d}
\end{center}
\end{figure}

NEMO-3 calorimeter modules~\cite{Arnold:2004xq} were manufactured to conform to the overall cylindrical geometry of the detector.  Each module faces the isotopic foil (a source of double beta transitions) and is composed of a scintillator block, a light guide, and a $\rm 3~inch$ or $\rm 5~inch$ PMT.  The detector is azimuthally divided into 20 identical wedge sectors, each assembled as a tracker and calorimeter with a source foil, as depicted  in Figure~\ref{figure:blocks3d}. The scintillators hermetically cover the two cylindrical walls which surround the foil and the tracking volume. There are three types of blocks comprising the walls and there is also partial scintillator coverage of the top and bottom end-caps.  On the interior wall, there are two IN blocks which are mirror-symmetric. On the exterior wall, there are two types of blocks: center (EC) and edge (EE) type blocks, with EE blocks symmetrically placed on either side of an EC block. The dimensions of these scintillator blocks are given in Table~\ref{table:blocks}. Simulations of the response of these three types of blocks was the primary objective of our work reported here.

The scintillator material is composed of polystyrene (PS) (98.49\% by weight), a primary dopant p-terphenyl (pTP) (1.5\% by weight) and a wavelength shifter 1,4-bis(5-phenyloxazol-2-yl) benzene (POPOP) (0.01\% by weight).  The light guides are made out of polymethylmethacrylate (PMMA) and serve as the interface between the PMTs and the scintillators.  Five layers of $\rm 70 ~\mu m$ thick Teflon ribbon are wrapped around the four side walls of each scintillator and two layers of single-sided $\rm 6 ~\mu m$ thick aluminized Mylar foil are wrapped around the entrance face.  One layer of double-sided $\rm 12 ~\mu m$ thick aluminized Mylar foil is wrapped around the other five faces to protect against light produced by Geiger discharge in the tracking region.  Bicron BC600 optical glue with a nominal thickness of $\rm 100 ~\mu m$ is used to couple the light guides to the scintillator blocks.

\begin{table}[htb]
\begin{center}
\caption{Types and dimensions of NEMO-3 inner and outer wall scintillator blocks. The blocks
were made at the Joint Institute for Nuclear Research (JINR) in Dubna, Russia.~\cite{Arnold:2004xq}}
\label{table:blocks}
\vskip0.05in
\newcommand{\m}{\hphantom{$-$}}
\renewcommand{\arraystretch}{1.2} 
\begin{tabular}{@{} | c | c | c | c |}
\hline
Block type     		& IN  			     & EC            	        & EE  	\\
\hline\hline
Thickness      		& 98 to 110 mm	& 99 mm	        & 99 to 123 mm 		\\ \hline
Height         		& 153 mm			& 200 mm		& 200 mm 			\\ \hline
Width          		& 138 to 154 mm	& 218 mm		& 218 to 230 mm 		\\ \hline
Associated PMT~\cite{Hamamatsu}	& R6091 (3 inch)       &  R6594 (5 inch)    	& R6594  (5 inch)     \\ \hline
Total number   		& 680            		& 260           	& 520                  		\\ \hline
\end{tabular}\\[2pt]
\end{center}
\end{table}

\subsection{Input components of the simulation model}

\begin{figure}
\begin{center}
\includegraphics[width=15pc]{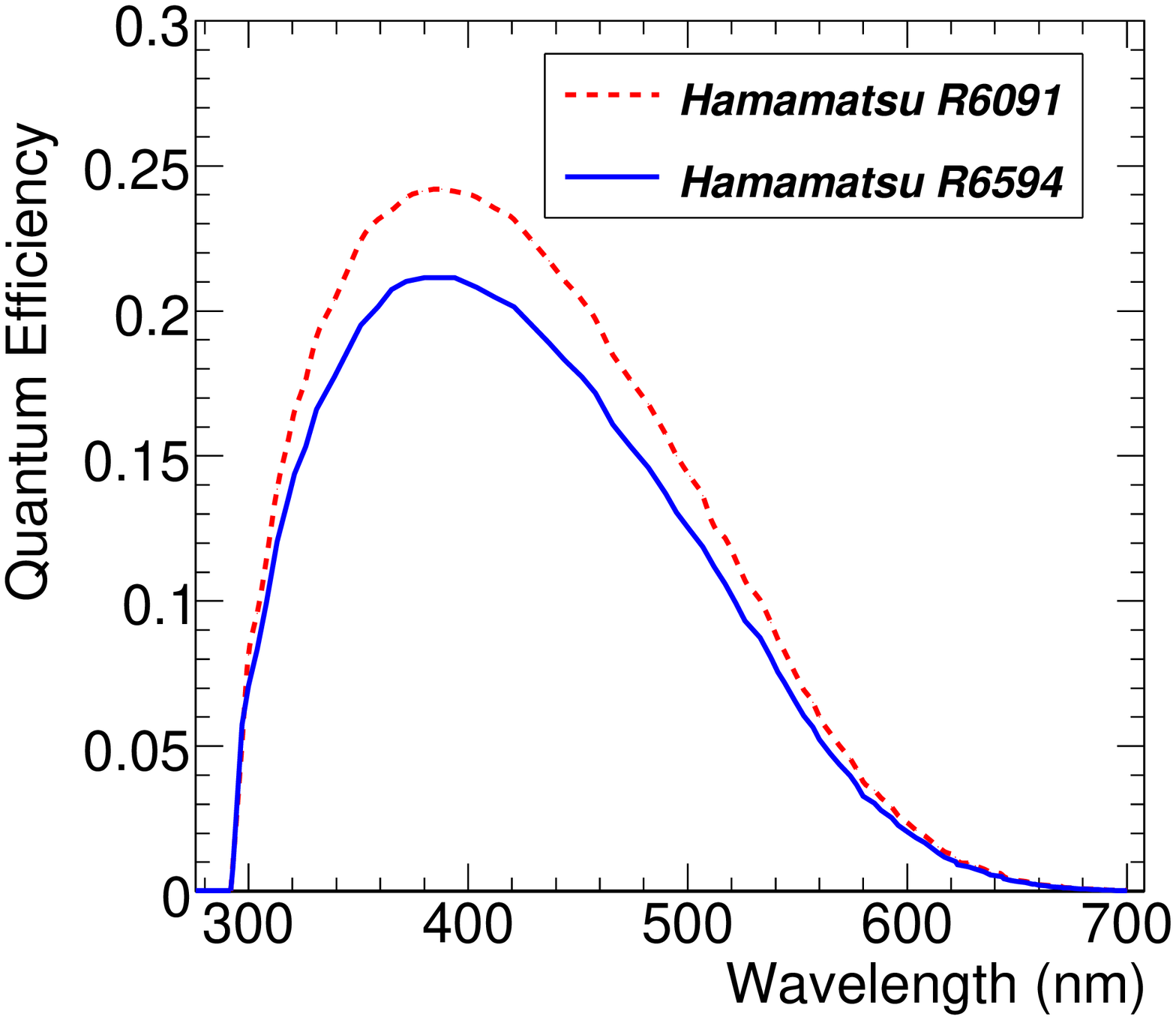}
\includegraphics[width=15pc]{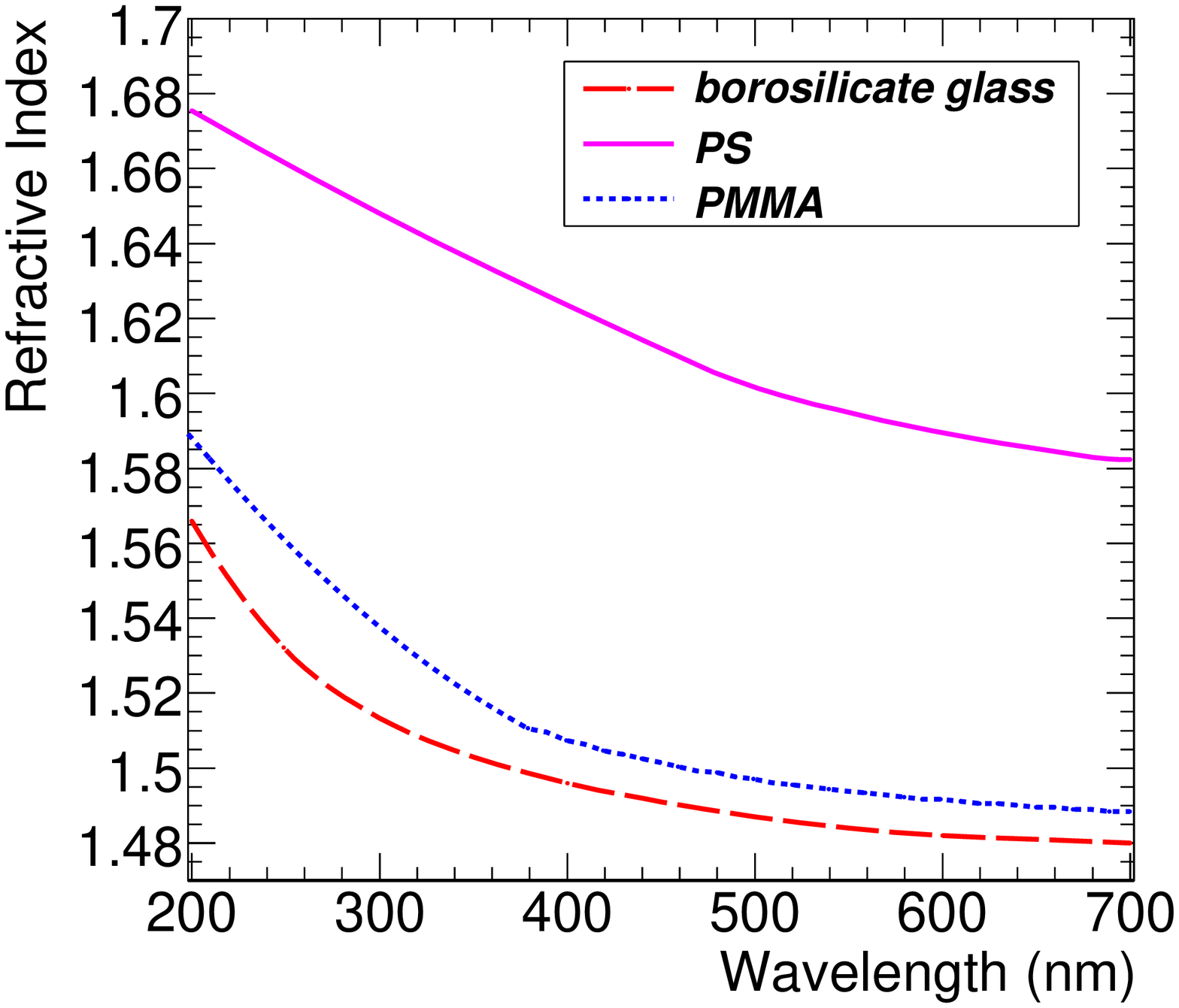}
\caption[Can also include note.]{Input data used in the NEMO-3 simulations.  Left: Typical quantum efficiency of Hamamatsu R6091 $\rm 3~inch$ and Hamamatsu R6594 $\rm 5~inch$ PMTs.  Right: Refractive indices of borosilicate glass, polystyrene, and PMMA.}
\label{fig:pmts_and_refractive_ind}
\end{center}
\end{figure}

We use spectral properties of all optical elements.  The effective quantum efficiency spectra\footnote{Effective quantum efficiency is the product of quantum efficiency and collection efficiency.} for the PMTs are shown in Figure~\ref{fig:pmts_and_refractive_ind}, where data are taken from~\cite{Hamamatsu}. The refractive index data for borosilicate glass are given by the Cauchy dispersion law $\rm n_{glass} = 1.472 + 3760/\lambda^2$, where $\lambda$ is the photon wavelength~\cite{Motta:2004yx}.  Figure~\ref{fig:pmts_and_refractive_ind} also shows the refractive index input data for the PMMA light guides and for scintillator polystyrene~\cite{Kasarova:2007}, which we have linearly extrapolated in the low wavelength ($\rm 200-300~nm$) region.  

\begin{figure}
\begin{center}
\includegraphics[width=15pc]{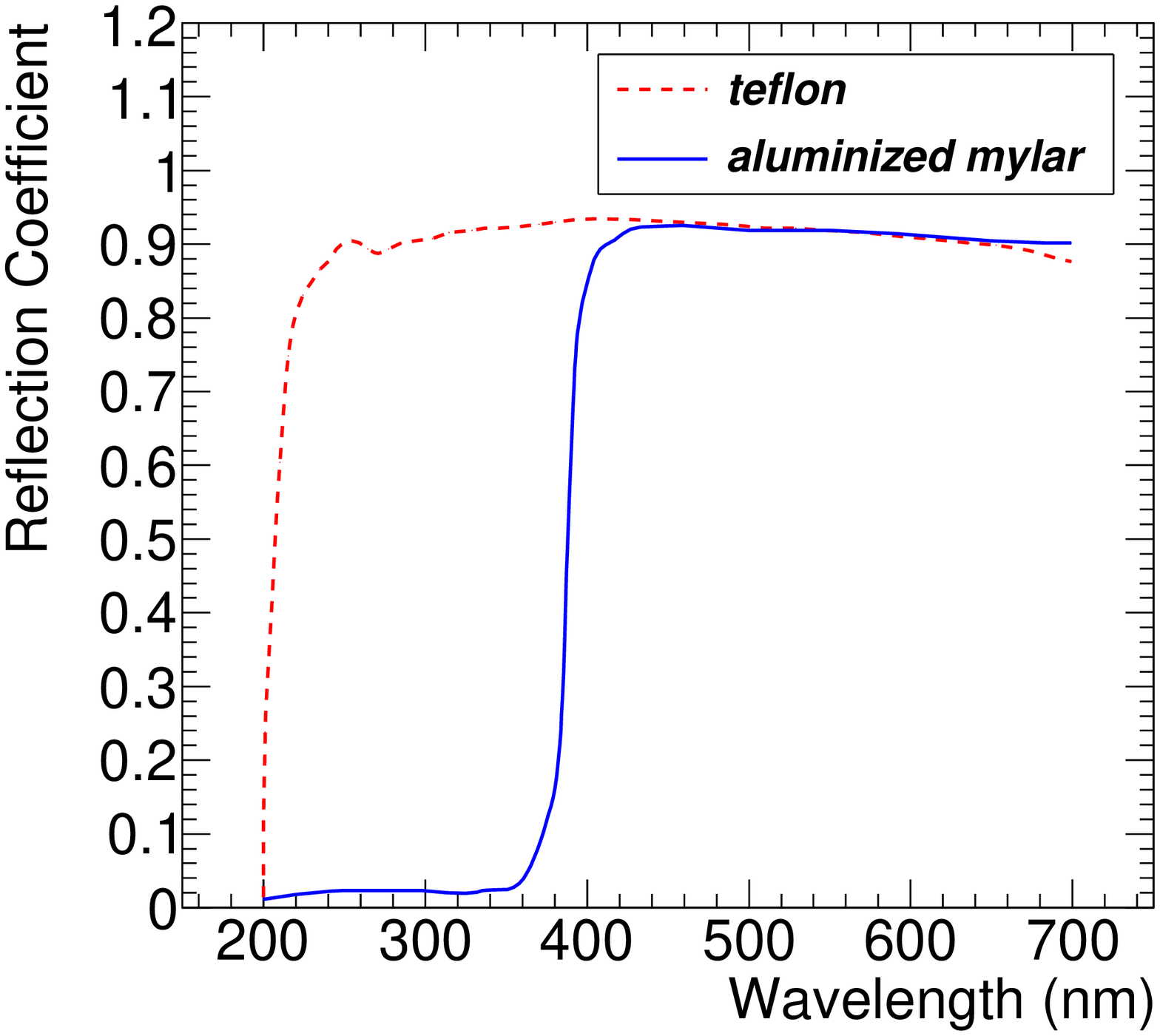}
\includegraphics[width=15pc]{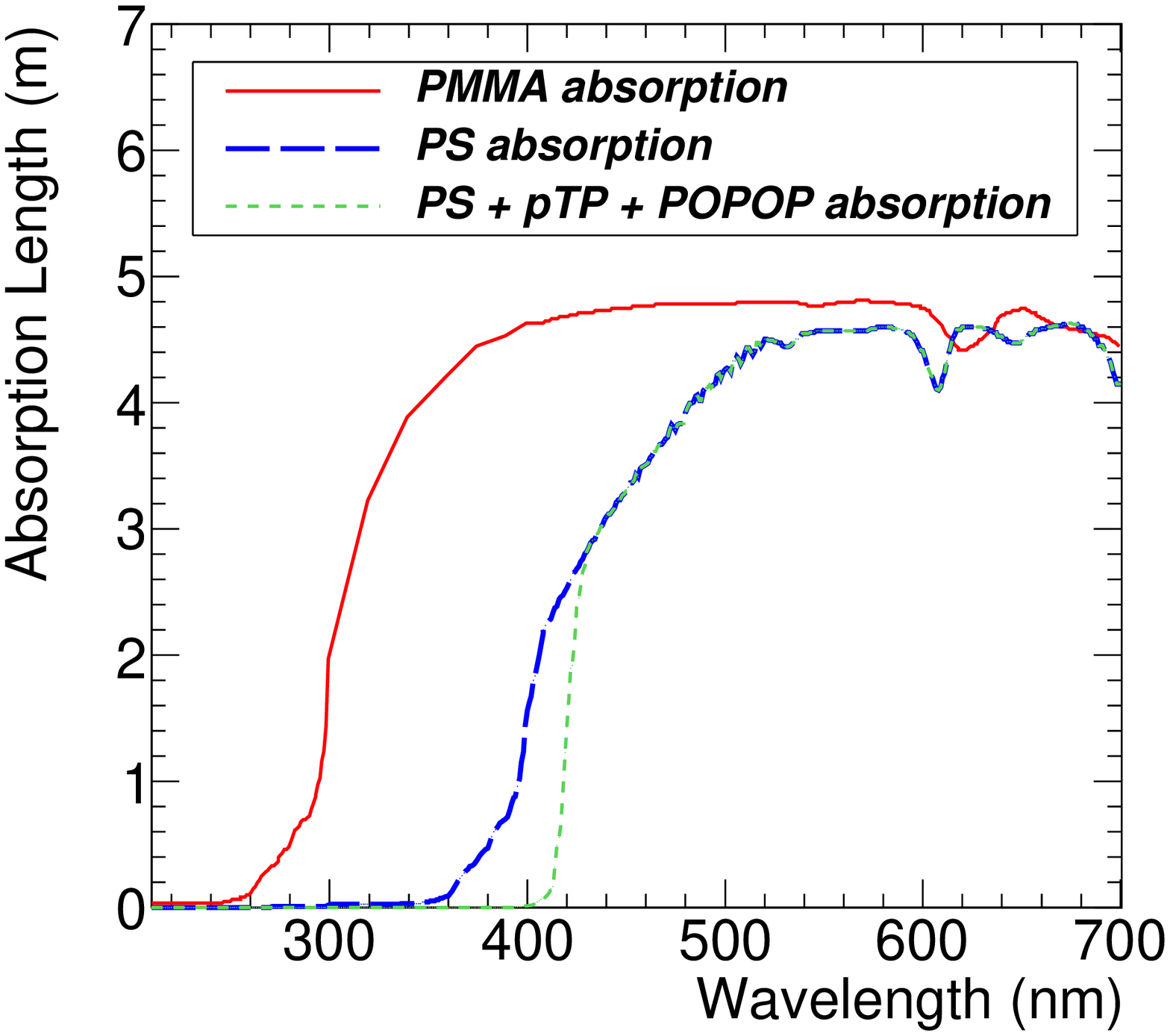}
\caption{Other input data used in the NEMO-3 simulations.  Left: Reflection coefficients for the Teflon tape and aluminized Mylar.  Right: Absorption lengths for PMMA, polystyrene, and the combined polystyrene/pTP/POPOP scintillator, where the absorption data were taken from~\cite{Senchishin:2006qw} and~\cite{Artikov:2005}.  The combination defines the scintillator absorption length after introduction of the fluors to the polystyrene. }
\label{fig:reflection_coefficients_and_absorption}
\end{center}
\end{figure}

The measured reflection coefficients of Teflon and aluminized Mylar used in the simulations are shown in Figure~\ref{fig:reflection_coefficients_and_absorption}.  We assume a $\rm 50 ~\mu m$ air gap between the scintillator block and the Teflon/Mylar wrappings whose reflectivities are modeled as 100\% Lambertian for Teflon and 100\%  specular for aluminized Mylar.  The scintillator blocks were treated under water with 1200 grit sandpaper to obtain diffusive reflection at the surfaces~\cite{Arnold:2004xq}.  To address this, we incorporate the surface roughness parameter ``sigma\_alpha'' of GEANT4 with a value of 360 degrees. 


\begin{figure}
\begin{center}
\includegraphics[width = 20pc]{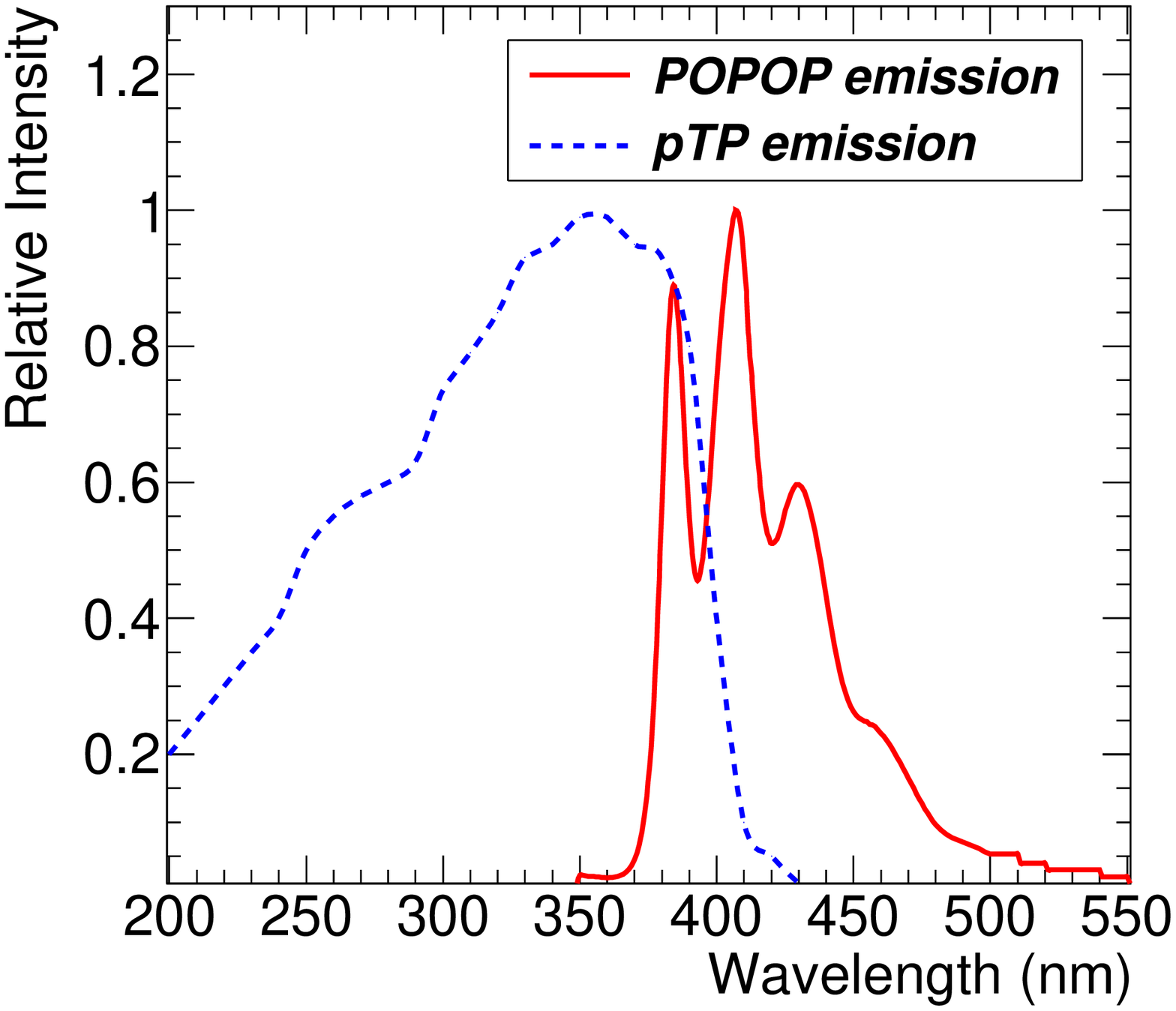}
\caption[Primary emission and absorption spectrum]{Primary emission spectrum for pTP and POPOP used in the simulations~\cite{Berlman:1971}.  }
\label{fig:primary_emission}
\end{center}
\end{figure}

We treat the polystyrene and the primary dopant, pTP as a single entity and define the primary emission spectrum to be that of the pTP alone, since the number of emitted photons from the polystyrene is negligible compared to the number of photons generated by pTP. The pTP emission spectrum is shown in Figure~\ref{fig:primary_emission}.  Photons emitted from pTP can either propagate in the polystyrene according to the bulk absorption length (BAL) of the polystyrene and pTP mixture or interact with a POPOP wavelength-shifting molecule.  Stokes shifting is determined by the combined absorption length of PS/pTP/POPOP with effective emission governed by the POPOP emission spectrum using data shown in Figures~\ref{fig:reflection_coefficients_and_absorption} and~\ref{fig:primary_emission}.  We account for the molecular quantum yield of POPOP by having absorbed photons re-emitted with 93\% probability~\cite{Berlman:1971} and at a wavelength equal to or greater than the absorbing wavelength,\footnote{A more precise treatment is possible for example, using Jablonsky diagrams for energy level spacing but would require an additional layer of complexity which we chose not to introduce at this time~\cite{Lakowicz}.} to obey energy conservation~\cite{Lakowicz}.  In GEANT4, two BALs must be specified: one for processes where the primary photon is absorbed in PS/pTP, and one for processes where the primary photon is absorbed by POPOP with the possibility of multiple wavelength shifts. 



\section{Energy calibration and impact corrections}

For energy calibration and alignment of blocks, the NEMO-3 experiment uses radioactive sources that are periodically inserted into the apparatus in well-defined positions.  Measurements are taken using sixty $\rm ^{207}Bi$ sources with a mean activity of approximately $\rm 210~Bq$.  Three sources per sector are used and each twenty-four hour run yields approximately two (three) thousand useful electron tracks for each IN (EE and EC) scintillator module.  The bismuth decay provides conversion electrons with energies of $\rm 482~keV$, $\rm 976~keV$, and $\rm 1682~keV$ (K-lines).  The electrons lose energy due to crossing several materials including the kapton window of the calibration tube \cite{Arnold:2004xq}, the helium gas, and the scintillator wrapping.  The mean energy losses are estimated to be $\rm 45~keV$ and $\rm 40~keV$ for the $\rm 482~keV$ and $\rm 976~keV$ electrons, respectively.  The relation between the PMT charge signal and the energy deposited in the scintillator block is linear from $\rm 150~keV$ to $\rm 4~MeV$.  We apply a linear formula to calculate the energy of each electron event: $E = \alpha(C-P) + \beta$ where $C$ is the ADC value of the scintillator, $P$ is the pedestal and $\alpha$ and $\beta$ are fit parameters.  The fit takes into account the energy loss calculated for each electron according to its energy and measured track length.  The energy resolution,$\rm R_{FWHM}$ at 1 MeV, for each block, is obtained from the width of the 976 keV peak assuming

\begin{equation}
\rm 
R_{FWHM}(E) = \frac{FWHM(E)}{E} = \sqrt{\frac{A}{E}} 
\label{eq:resolution}
\end{equation}
\noindent
where $\rm A(MeV)$ is a constant with a range from 0.014 to 0.032 for all blocks. The mean value of $\rm R_{FWHM}$ at 1 MeV for EC blocks is $\rm 13.8\%$, for EE blocks is $\rm 13.5\%$, and for IN blocks is $\rm 16.7\%$, as summarized in Table~\ref{table:eres_results}.

The response of each scintillator block depends upon the entrance point of the electron. NEMO-3 data show a weak dependence of 1 to 2\% for blocks equipped with $\rm 3~inch$ PMTs and a stronger dependence of up to 10\% for larger blocks equipped with $\rm 5~inch$ PMTs.  This effect has a non-negligible consequence on the energy resolution.  The ADC value and energy loss were measured for each electron in each scintillator block to correct for this dependency.  The front faces of the scintillator blocks were divided into nine ($\rm 3 \times 3$) rectangles of equal areas for modules equipped with $\rm 3~inch$ PMTs and twenty-five ($\rm 5 \times 5$) rectangles of equal areas for modules equipped with $\rm 5~inch$ PMTs.  ADC histograms were obtained for each impact region.  Then the energy distribution around the $\rm 976~keV$ electron peak was refit to obtain the parameter $\rm A$ of the resolution function.  The mean energy resolutions $\rm R_{FWHM}(E)$ at $\rm 1~MeV$ calculated using these corrections are given in Table~\ref{table:eres_results}.


\section{Results of simulations}

The goal of our simulations was to understand the measured energy resolution and the response non-uniformity as a function of impact position. We investigated the light collection from electrons incident on the faces of three types of NEMO-3 scintillator blocks.  In our model, 1 MeV electrons were generated $\rm 60~cm$ away from each type of scintillator block in vacuum, for direct comparison with the NEMO-3 geometry and the measured energy resolution.  The angular distribution of simulated electrons matched the solid angle of the region of interest on the block surface.  The block faces were divided in the same fashion as the measured data.  We simulated 2,500 electrons with energy of $\rm 1~MeV$ in each region of the grid.  Following the literature, we assumed the light yield for the polystyrene-based NEMO-3 blocks to be $\rm 8,000$ photons per MeV for electrons~\cite{Senchishin:2006qw}.  This value is not well known for our scintillator and is a source of systematic uncertainty.   The simulated energy resolutions in three types of blocks are compared to measured values shown in Table~\ref{table:eres_results}.  The energy resolution $\rm R_{FWHM}$ is calculated as $\rm R_{FWHM}={2.35}/{\sqrt{N_{pe}}}$, where $\rm N_{pe}$ is the number of photoelectrons registered by the PMT per each simulated electron.

\begin{table}
\begin{center}
\caption{Simulated and measured mean energy resolutions at 1~MeV for three different types of NEMO-3 scintillator modules.  Simulation uncertainty is taken from Table~\ref{table:uncertainty_results} for $10$\% input variations.  Measurement uncertainty is calculated from the collective response variability between individual blocks. }
\label{table:eres_results}
\vskip0.05in
\newcommand{\m}{\hphantom{$-$}}
\renewcommand{\arraystretch}{1.2} 
\begin{tabular}{@{} | c | c | c | c |}
\hline
Block Type    	        & Simulated $\rm R_{FWHM}$ & Measured $\rm R_{FWHM}$ 	& Measured $\rm R_{FWHM}$ \\
      			&                          & with impact corrections    & without impact corrections	\\
\hline\hline
EC        		& 14.4 $\pm$ 1.1    	& 13.8 $\pm$ 0.3    	        & 15.6 $\pm$ 0.3      \\ \hline
EE        		& 14.0 $\pm$ 1.1    	& 13.5 $\pm$ 0.2   	        & 14.9 $\pm$ 0.2      \\ \hline
IN        		& 14.9 $\pm$ 1.1    	& 16.7 $\pm$ 0.2   	        & 16.8 $\pm$ 0.2      \\ \hline
\end{tabular}\\[2pt]
\end{center}
\end{table}

\subsection{Spatial non-uniformity of the response}

\begin{figure}[t]
\begin{center}
\includegraphics[width=22pc]{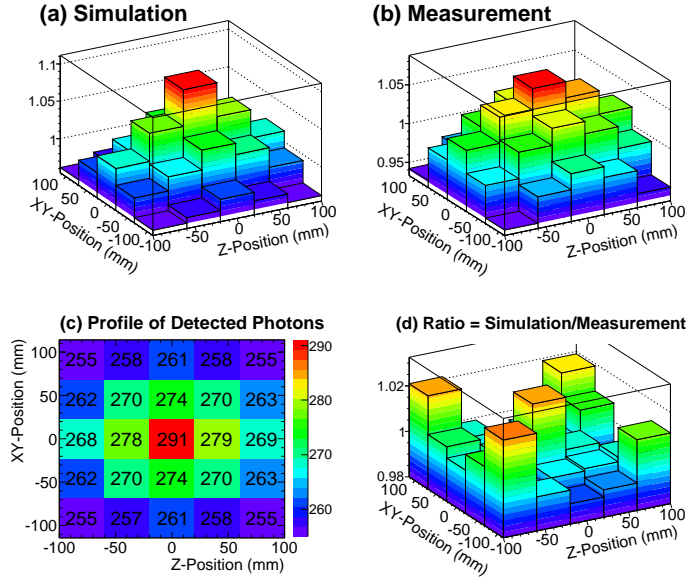}
\caption{Results from $\rm 1~MeV$ electrons incident on the EC block. Simulated (a) and measured (b) response, normalized to the mean response.  (c) The mean number of photo-electrons collected in each sub-region. (d) The ratio of simulation to measurement.}
\label{fig:ec_mapping}
\end{center}
\end{figure}

\begin{figure}[t]
\begin{center}
\includegraphics[width=22pc]{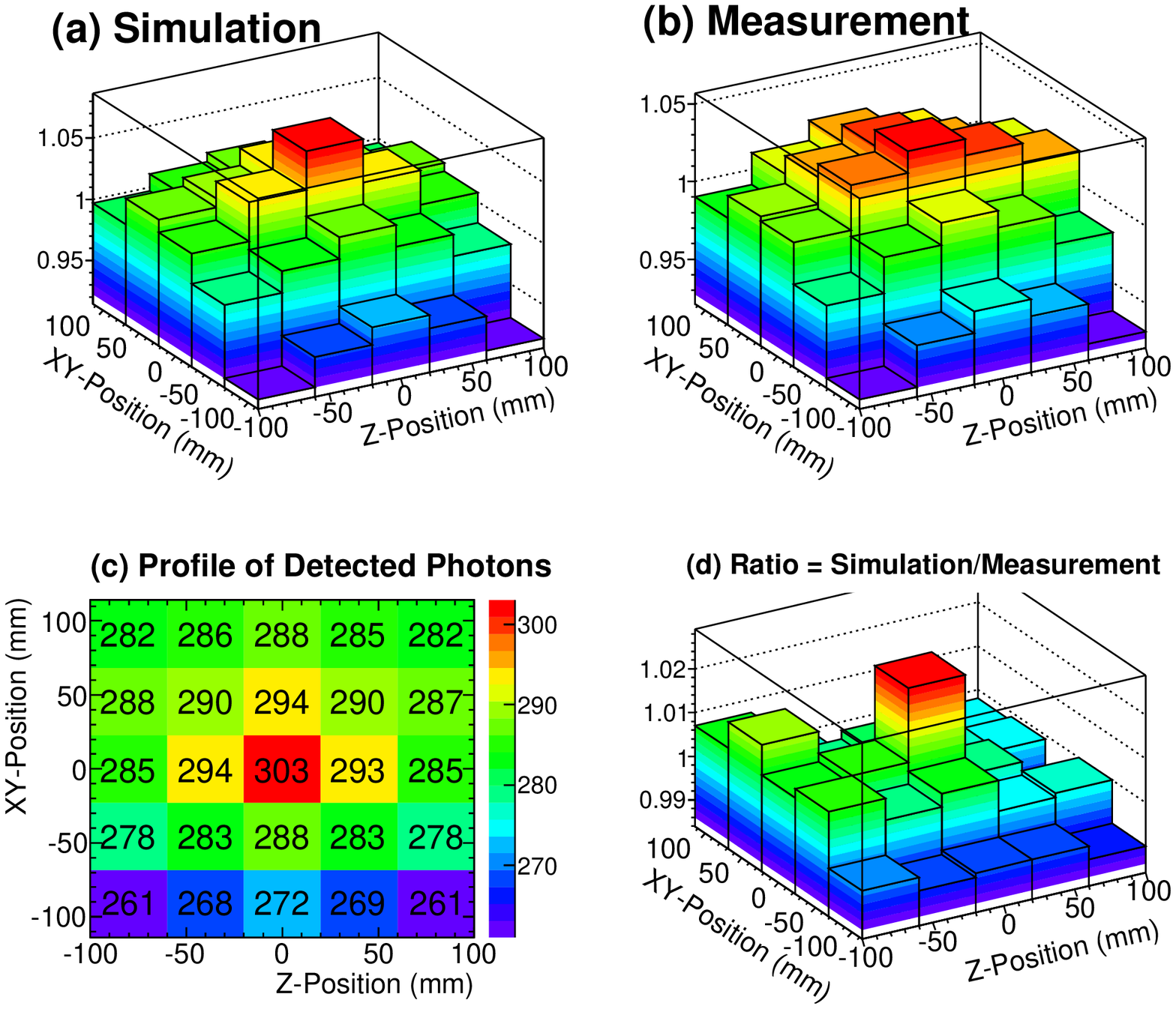}
\caption{Results from $\rm 1~MeV$ electrons incident on the EE block. Simulated (a) and measured (b) response, normalized to the mean response.  (c) The mean number of photo-electrons collected in each sub-region. (d) The ratio of simulation to measurement.}
\label{fig:ee_mapping}
\end{center}
\end{figure}

\begin{figure}[t]
\begin{center}
\includegraphics[width=22pc]{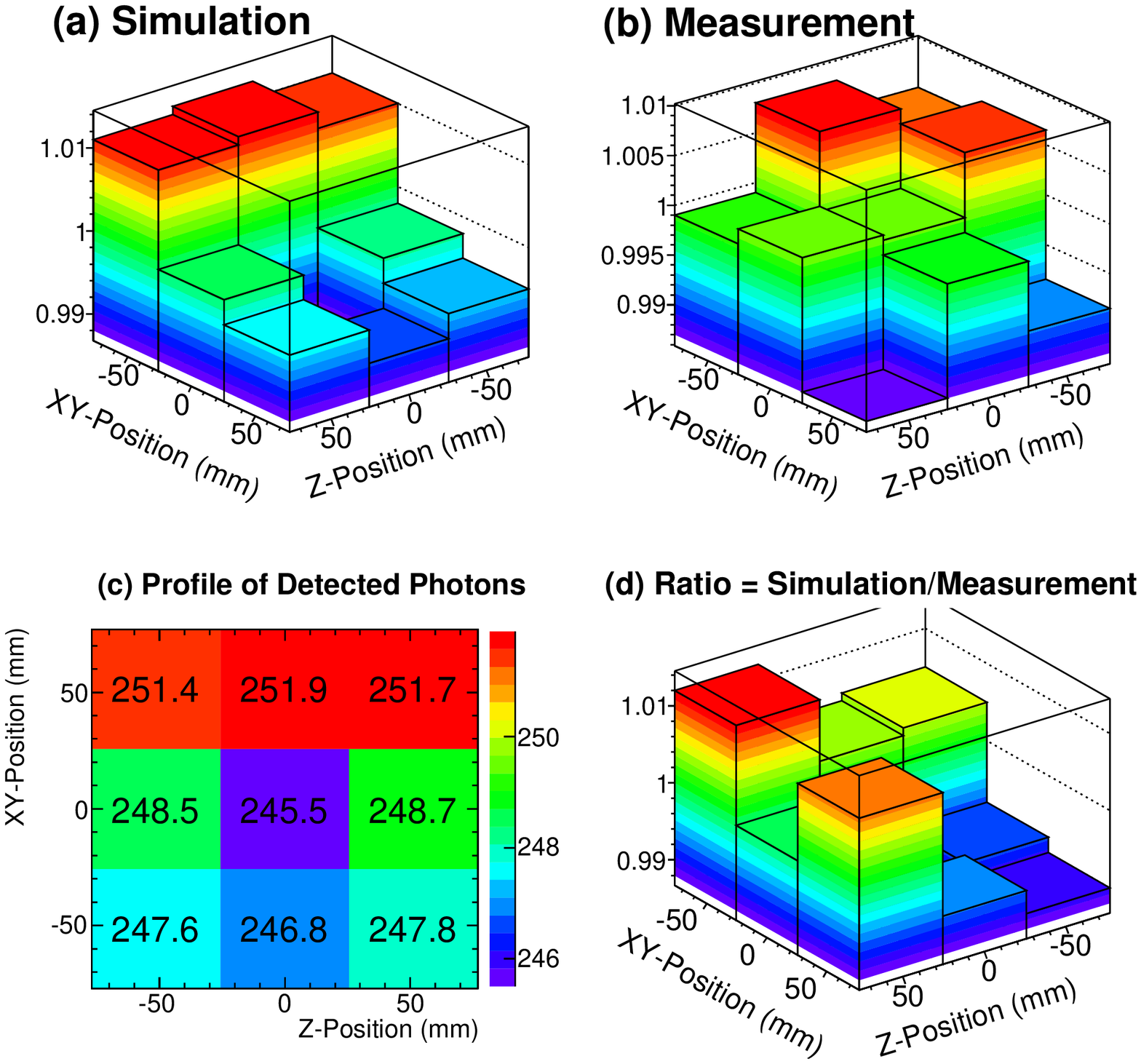}
\caption{Results from $\rm 1~MeV$ electrons incident on the IN block. Simulated (a) and measured (b) response, normalized to the mean response.  (c) The mean number of photo-electrons collected in each sub-region. (d) The ratio of simulation to measurement.}
\label{fig:in_mapping}
\end{center}
\end{figure}

We compared the simulated and measured spatial response non-uniformity by normalizing the number of simulated photoelectrons and the measured PMT charge from each region to the average over the entire block, shown in Table~\ref{table:eres_results}.  The EC block response uniformity is shown in Figure~\ref{fig:ec_mapping}(a) and (b).  The response reflects the symmetry of the block itself.  For the EE block, one side is deeper than the other which results in a skewed response as is clearly seen in Figure~\ref{fig:ee_mapping}(a) and (b).  The smaller IN block displays better uniformity although, one can still clearly identify the structural shape of the block in the results shown in Figure~\ref{fig:in_mapping}(a) and (b). Our simulations reproduce the measured response non-uniformity of each NEMO-3 block type.  The number of simulated photoelectrons collected by each type of module is shown in Figures~\ref{fig:ec_mapping}(c), \ref{fig:ee_mapping}(c), and \ref{fig:in_mapping}(c). The ratio of the normalized simulation value to the normalized measured value show variations no greater than 2\% for all block types, as shown in Figures~\ref{fig:ec_mapping}(d), \ref{fig:ee_mapping}(d), and \ref{fig:in_mapping}(d).  The ratio of the minimum to maximum response in collected photoelectrons is 86\% and 92\% for EC and EE simulations, respectively, and $\rm 88$\% for measurement in both block types.  The ratio of the minimum to maximum response is $\rm 97$\% for IN block simulations and $\rm 98$\% for measurement.  The statistical uncertainty in the mean number of photoelectrons collected for each bin is 0.04\%.


\subsection{Characteristics of simulated photons}

\begin{figure}[htb]
\begin{center}
\includegraphics[width=12.2pc]{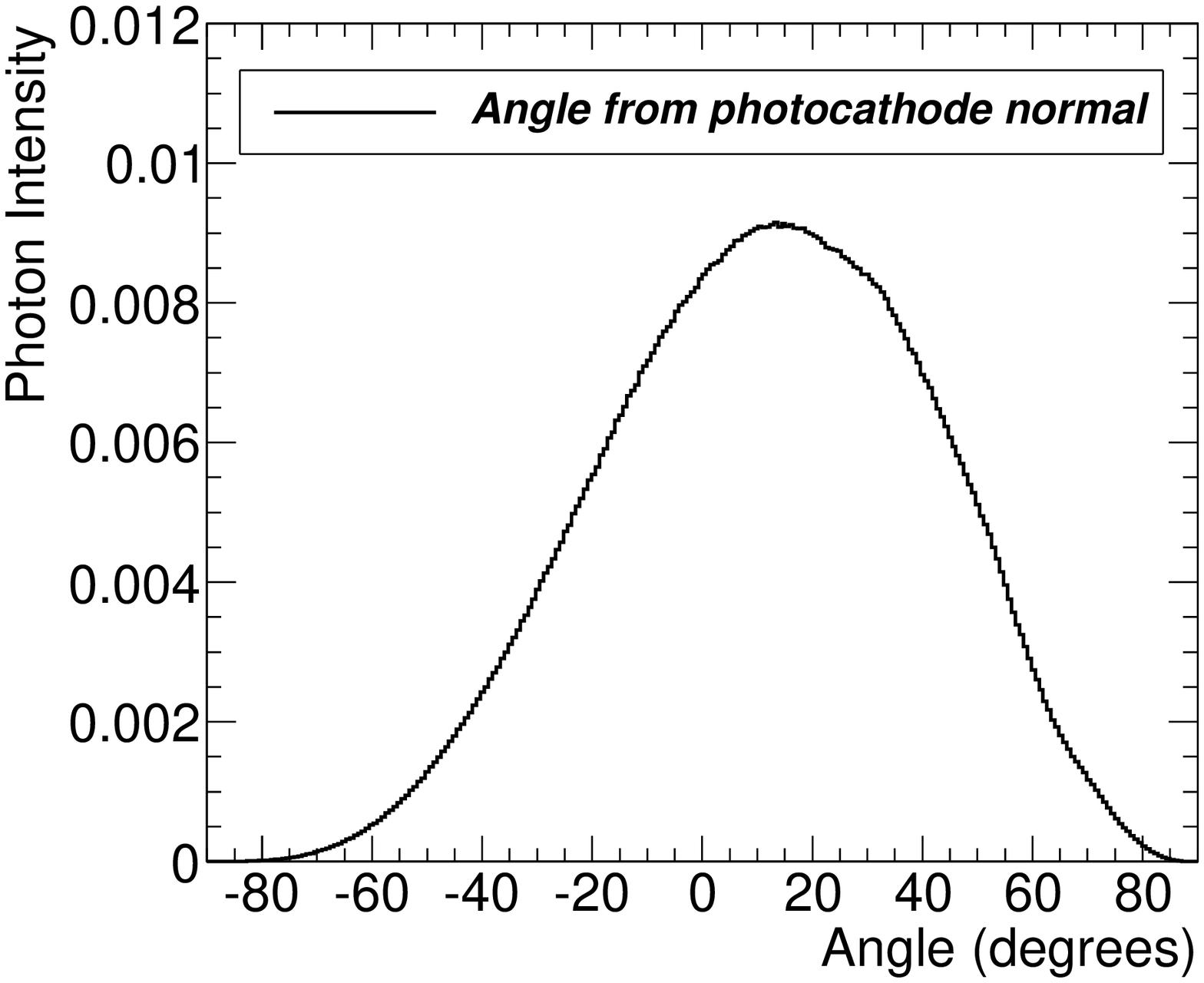}
\includegraphics[width=12.2pc]{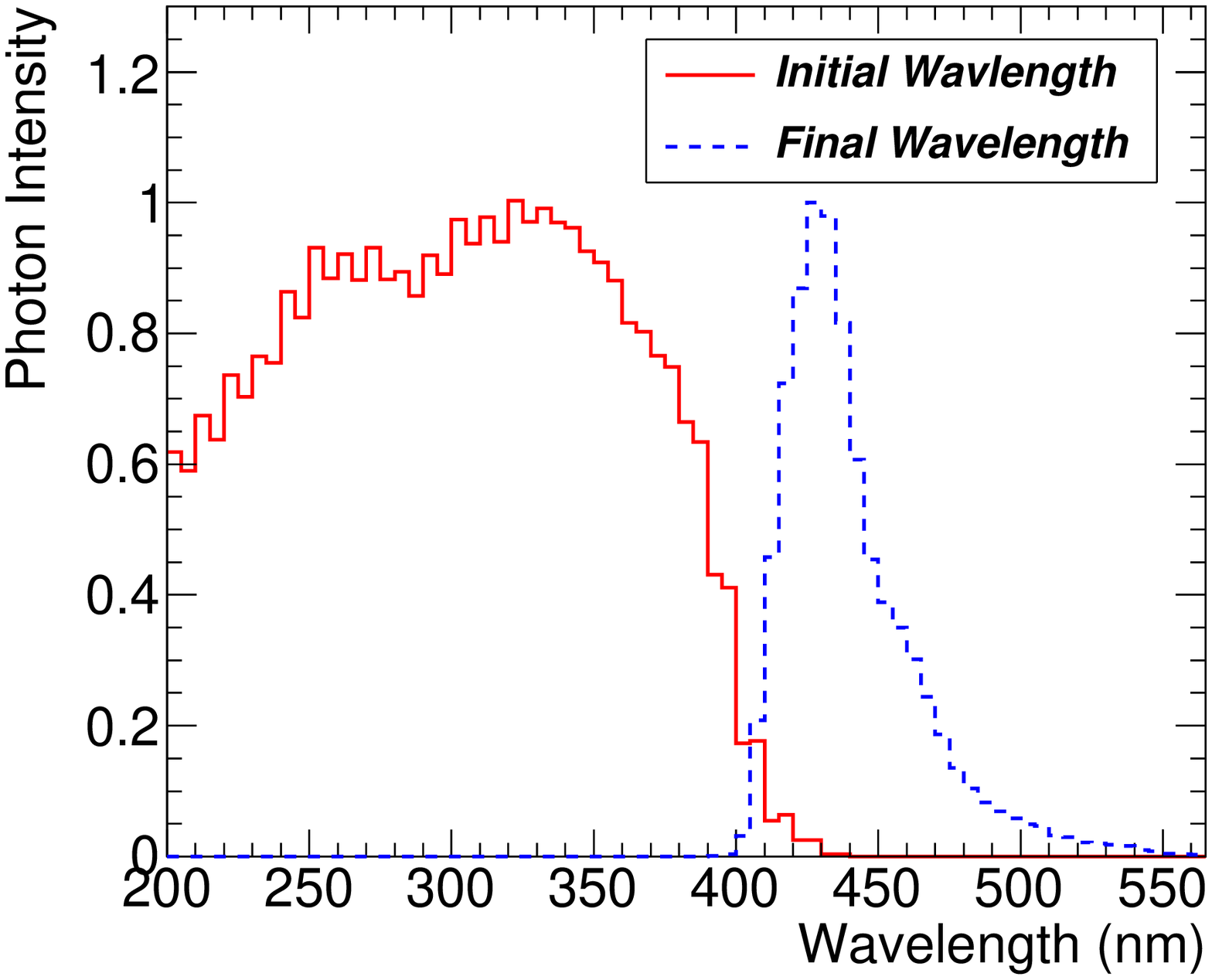}
\includegraphics[width=12.2pc]{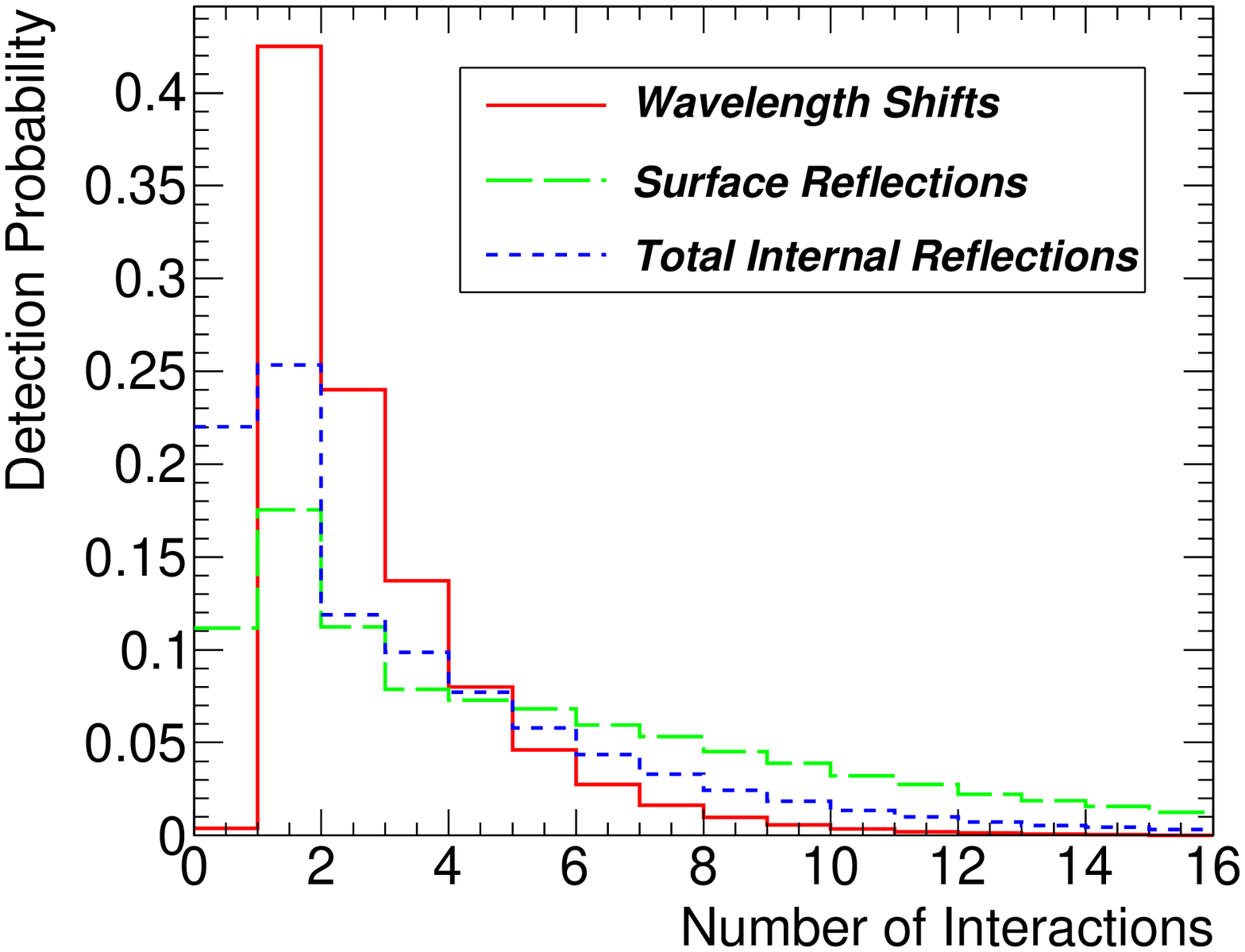}
\caption[Initial Wavlength]{Left: Incident angle of detected photons relative to the photocathode normal for the EC block.  Center: Initial and final wavelengths of detected photons simulated for the EC block.  The initial wavelength distribution is primarily that of the pTP emission.  The final wavelength distribution is reminiscent of the POPOP emission spectrum with degradation due to absorption at short wavelengths and poor PMT quantum efficiency at long wavelengths.  Right:  The detection probability per photon as a function of the number of wavelength shifts, number of reflections off the wrapping surfaces, and number of total internal reflections.}
\label{fig:initial-final-wavelength}
\end{center}
\end{figure}

Figure~\ref{fig:initial-final-wavelength} shows the incident angle of detected photons measured from the normal to the photocathode surface for the EC block.  Figure~\ref{fig:initial-final-wavelength} also shows the initial and final wavelength profiles of the detected photons for the EC block.  The distribution resembles that of the POPOP emission spectrum.  Shorter wavelength photons are absorbed in the bulk scintillator material and on the surface of the wrappings resulting in a suppression below $\rm 400~nm$. Longer wavelength photons are supressed due to the low PMT quantum efficiency at longer wavelengths.  Figure~\ref{fig:initial-final-wavelength} also shows the detection probability as a function of the number of wavelength shifts per photon, the number of reflections from the wrapping surfaces, and the number of total internal reflections.  Each individual photon, for example, can undergo several wavelength shifts and several reflections (specular, diffuse, and total internal) before reaching the PMT.  The distribution of wavelength shifting events suggests that a large fraction of the photons from pTP have been wavelength shifted. However, the probability for zero wavelength shifts is finite because pTP emission can take place above 400 nm, out of the range of large bulk attenuation.


\subsection{Dependence of simulations on input parameters}

For the EC block, we have varied the simulation input parameters to evaluate the dependence of the energy resolution on these changes. The PMT quantum efficiency, the absorption length of the scintillator, the reflectivity of Teflon and Mylar wrappings, and the light yield were decreased and increased by $5\%$ and $10\%$. These fractional changes reasonably reflect physical variations in production or manufacturing of most components. We then compared the result of each change to the central value of the energy resolution of 14.4\%.  Table~\ref{table:uncertainty_results} summarizes our studies. 

\begin{table}[h]
\begin{center}
\caption{Changes of the EC block energy resolution at 1~MeV due to variations 
of input parameters. We increased and decreased the parameter values by 
$5\%$ and $10\%$ and quote the difference in the energy resolution between the central
value of 14.4\% and an average of the change.
Total change is calculated by adding the individual changes in quadrature.  }
\label{table:uncertainty_results}
\vskip0.05in
\newcommand{\m}{\hphantom{$-$}}
\renewcommand{\arraystretch}{1.2} 
\begin{tabular}{@{} | l | c | c |  }
\hline
Optical Parameter    	&  5\%  		& 10\%      \\ 
\hline\hline
Quantum Efficiency      	& 0.34\% 		& 0.75\%      	\\ \hline
Absorption Length        	& 0.12\% 		& 0.31\%      	\\ \hline
Teflon Reflectivity      	& 0.92\%		& 2.50\%       	\\ \hline
Mylar Reflectivity       	& 0.54\% 		& 0.98\%      	\\ \hline
Light yield			& 0.30\%		& 0.75\%	\\ \hline
\hline
Total change        		& 1.17\% 		& 2.90\%      	\\ \hline
\end{tabular}
\end{center}
\end{table}

\subsection{Dependence of energy resolution on spectral properties}

We evaluated the importance of using spectral properties of materials in our simulations of energy resolution of the EC block. We began with a simplified simulation of optical materials with wavelength-independent values and investigated the effect of systematically introducing the wavelength dependence into the simulation.  Initially, we fixed the value for the quantum efficiency at 21\%, the absorption length at $\rm 4.5~m$, the reflectivity for the Teflon and Mylar at 93\%, and the refractive indices at 1.5.  The wavelength-dependent quantum efficiency of the PMT was introduced first.  We then introduce a wavelength-dependent absorption length for the scintillator with no wavelength shifting and subsequently introduce wavelength-shifting dependence and molecular quantum efficiency.  Finally, we introduced wavelength-dependence for the Teflon and Mylar reflection coefficients, and the scintillator and borosilicate glass refractive indices.  The values here were chosen as best-guess estimates at a peak wavelength of 420 nm that one may assume for a monochromatic simulation.  The results for each step are shown in Table~\ref{table:systematics_results}. The complete model reproduces the measured energy resolution within systematic uncertainty.   It has been suggested that there is a variation in the PMT quantum efficiency as a function of the incident angle of the photon~\cite{Motta:2004yx}.  If we include this effect, assuming it is also relevant for NEMO-3 PMTs, the energy resolution for the complete model improves by about 0.3\% while the spatial response distribution is left unchanged.

\begin{table}[h]
\begin{center}
\caption{Simulated FWHM of energy resolution at 1~MeV of EC scintillator blocks in NEMO-3  for increasingly comprehensive model parameters. The initial values are best-guess estimates at a peak wavelength of 420 nm that one may assume for a simplified monochromatic simulation.  Wavelength-dependence is introduced systematically to show the effect.}
\label{table:systematics_results}
\vskip0.05in
\newcommand{\m}{\hphantom{$-$}}
\renewcommand{\arraystretch}{1.2} 
\begin{tabular}{@{} | l | c |  }
\hline
                                                   				&   $\rm R_{FWHM}$  \\ 
All quantities fixed  							& \\
~~~~~quantum efficiency : 21\%  				& \\
~~~~~absorption length : 4.5 m 				& \\
~~~~~Teflon/Mylar reflectivity : 93\% 			& \\
~~~~~refractive indices : 1.5                            		&   11.2    	\\ \hline\hline
After introduction of $\lambda$ dependence to: 	&         	\\ \hline
Quantum efficiency (at constant abs. length)		&   11.6 	\\ \hline
Absorption length                                  			&   13.8    \\ \hline
Absorption length with Stokes shifting             		&   14.2    \\ \hline
Mylar reflection coefficient                       			&   14.4    \\ \hline
Teflon reflection coefficient                      			&   14.4    \\ \hline
All refractive indices                             			&   14.4    \\ \hline
\hline
Complete model               					&   14.4    \\ \hline
Measured                     						&   13.8    \\ \hline
\end{tabular}\\[2pt]
\end{center}
\end{table}
\section{Modeling of SuperNEMO scintillator blocks }

We have used our simulations to facilitate a scintillator choice for the SuperNEMO experiment~\cite{Chauveau:2009zz}. SuperNEMO further plans to exploit the NEMO-3 technique of tracking and calorimetry. The new modular detector would incorporate about $\rm 100~kg$ of $\rm ^{82}$Se, $\rm ^{150}$Nd, or $\rm ^{48}$Ca, to reach neutrinoless double beta decay half-life sensitivity of about $\rm 1.5 \times 10^{26}$ years.  This goal requires that the new experiment significantly improves its energy resolution with respect to NEMO-3. We have been conducting R\&D and the developed simulation code is an important aid in our studies of the choice of materials, block shape and size, wrapping, and light collection.

\begin{figure}
\begin{center}
\includegraphics[width=20pc]{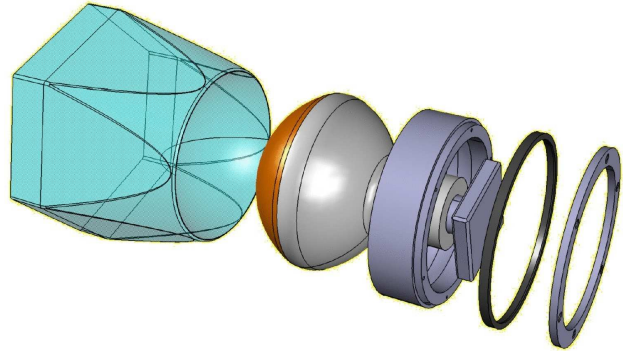}
\caption[Total Reflections]
{A drawing of a provisional SuperNEMO scintillator module showing the scintillator, PMT, and mounting brackets.   }
\label{fig:SuperNEMO_drawing}
\end{center}
\end{figure}

\begin{figure}
\begin{center}
\includegraphics[width=20pc]{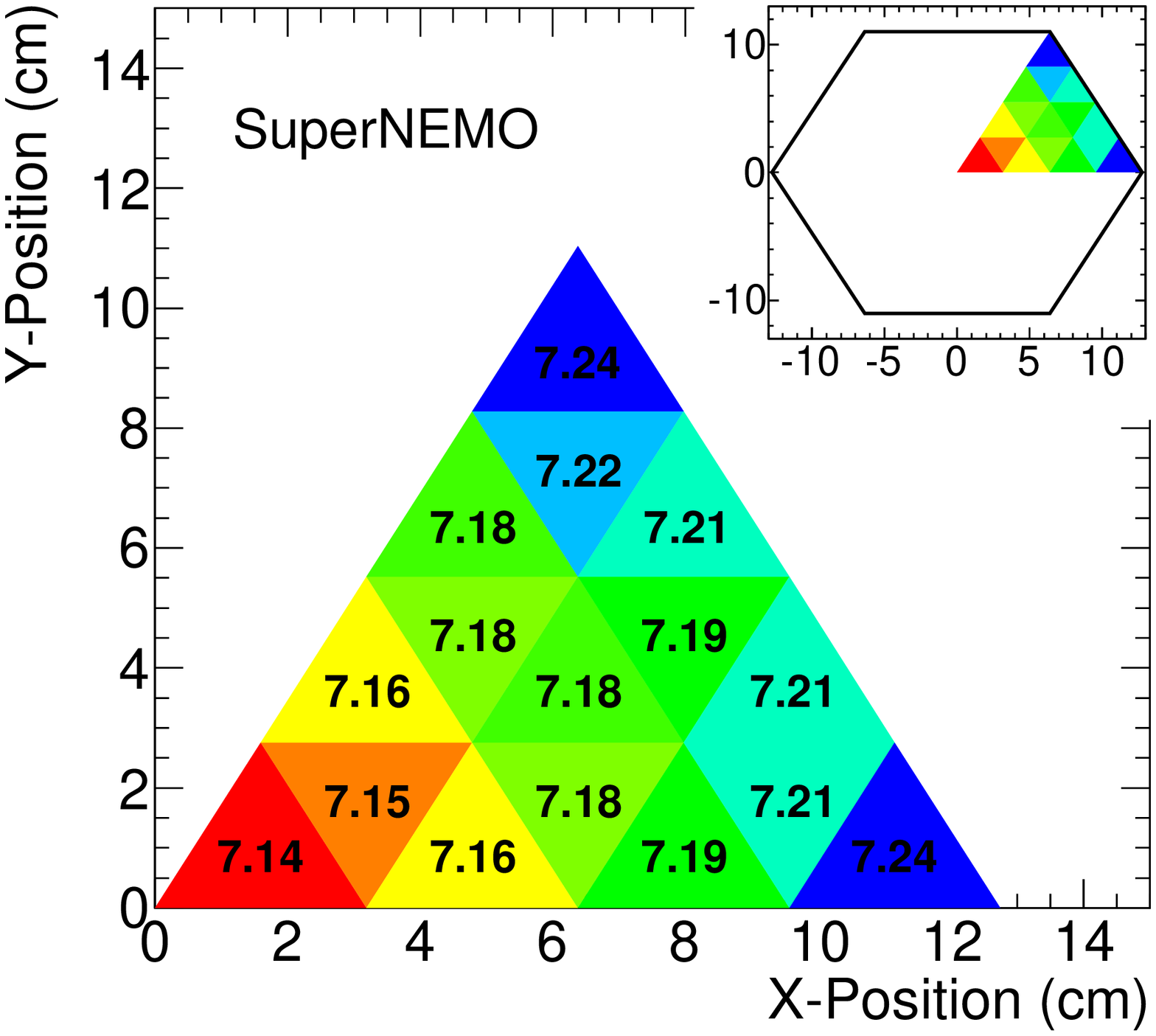}
\caption[Total Reflections]
{Spatial dependence of energy resolution (FWHM at 1~MeV) for one sixth of a SuperNEMO scintillator block.  }
\label{fig:SuperNEMO_resolution}
\end{center}
\end{figure}

A baseline provisional design for the SuperNEMO calorimeter calls for a hexagonal block with a proposed circumscribed radius of $\rm 22.5~cm$ made out of polyvinyltoluene (PVT) scintillator (e.g., Eljen EJ-200~\cite{Eljen}) coupled to a  super-bialkali $\rm 8~inch$ hemispherical PMT (e.g., Hamamatsu R5912-MOD). Figure~\ref{fig:SuperNEMO_drawing} shows a drawing of the proposed SuperNEMO scintillator module including the scintillator, PMT, and mounting brackets. The light yield of the PVT scintillator is taken to be 10,000 photons per MeV~\cite{Eljen}\footnote{This is compared to a polystyrene scintillator with nominal light yield of 8,000 photons per MeV}.  The quantum efficiency of the PMT was taken to be approximately 33\% at $\rm 420~nm$~\cite{Hamamatsu}\footnote{This is compared to a 5 inch NEMO-3 PMT with a quantum efficiency of 25\% at $\rm 420~nm$}.  We propose a high reflectivity aluminized Mylar from ReflecTech~\cite{ReflecTech} around the sides and entrance face of the scintillator block and a Teflon wrapping on the top face near the PMT.  Similar to NEMO-3, we assume an air gap of $\rm 50~\mu m$ between all block faces and the Mylar wrapping.  Our simulations show that with this configuration, assuming NEMO-3 scintillator absorption and emission, a resolution of $\rm 7.5 \pm 0.5\%$ (FWHM) at $\rm 1~MeV$~\cite{Pahlka:2008dw} is expected.  Using the symmetry of the block, each sixth of the hexagonal face is divided into 16 regions.  As shown in Figure~\ref{fig:SuperNEMO_resolution}, the energy resolution is fairly uniform across the face of the block. The mean energy resolution is 7.19\% (FWHM) at $\rm 1~MeV$ and the minimum and maximum is 7.14\% and 7.24\%, respectively.   Recently conducted preliminary measurements confirm our predictions~\cite{Chauveau:2009zz}.

\clearpage
\section{Conclusions}

We have constructed a GEANT4-based spectral model of NEMO-3 and SuperNEMO calorimeter blocks and compared the results of simulations with measurements of single electrons from $\rm ^{207}Bi$ sources.  The measured energy resolution and spatial dependence for the blocks were demonstrated to be in good agreement with simulations.  The EC blocks show a doubly-symmetric distribution of response about the center of the block while the EE blocks show a symmetric distribution about the z-axis, both in agreement with observations.  For the SuperNEMO block, the main elements improving the resolution over the NEMO-3 blocks are the 25\% increase in scintillator light yield, the 35\% increase in the PMT quantum efficiency, the larger PMT size, and the incorporation of the high reflectivity aluminized Mylar.  Additionally, by directly coupling the scintillator and PMT, we render light guides unnecessary which further improves the transparency and response uniformity.  

As expected, including the spectral properties of all materials and incorporating wavelength-shifting absorption and emission with the effects of fluorescent quantum yield in the scintillator improves the agreement of modeling with measurements. Our approach is necessary for a detailed understanding of high resolution plastic calorimeters. The importance of these spectral simulations increases with distances travelled by photons so such models are necessary to study large-scale scintillator detectors.

Acknowledgments:
We thank the staff at the Modane Underground Laboratory for their technical assistance in running the NEMO-3 experiment. We acknowledge support by the Grants Agencies of France, the Czech Republic, RFBR (Russia), STFC (UK), NSF, DOE, and DOD (USA).



\end{document}